\documentclass{article}
\usepackage{auxiliar/icmc2025_paper_template}
\usepackage{times}
\usepackage{ifpdf}
\usepackage{soul}
\usepackage{amsfonts}
\usepackage{amsmath}
\usepackage[english]{babel}
\usepackage{float}

\def\papertitle{Fractional Fourier Sound Synthesis}
\def\firstauthor{Esteban Gutiérrez}
\def\secondauthor{Rodrigo F. Cádiz}
\def\thirdauthor{Carlos Sing Long}
\def\fourthauthor{Frederic Font}
\def\fifthauthor{Xavier Serra}

\newif\ifpdf
\ifx\pdfoutput\relax
\else
   \ifcase\pdfoutput
      \pdffalse
   \else
      \pdftrue
  \fi
\fi

\ifpdf 
  \usepackage[pdftex,
    pdftitle={\papertitle},
    pdfauthor={\firstauthor, \secondauthor, \thirdauthor},
    bookmarksnumbered, 
    pdfstartview=XYZ 
   ]{hyperref}

  \usepackage[pdftex]{graphicx}
  \graphicspath{{./figures/}}
  \DeclareGraphicsExtensions{.pdf,.jpeg,.png}

  \usepackage[figure,table]{hypcap}

\else 
  \usepackage[dvips,
    bookmarksnumbered, 
    pdfstartview=XYZ 
  ]{hyperref}  

  \usepackage[dvips]{epsfig,graphicx}
  \graphicspath{{./figures/}}
  \DeclareGraphicsExtensions{.eps}

  \usepackage[figure,table]{hypcap}
\fi

\hypersetup{
    colorlinks,%
    citecolor=black,%
    filecolor=black,%
    linkcolor=black,%
    urlcolor=black
}

\title{\papertitle}

\fiveauthors
  {0.5in}
  {\firstauthor} {Music Technology Group\\Universitat Pompeu Fabra \\ %
    {\tt \href{mailto:esteban.gutierrezc@upf.edu}{esteban.gutierrezc@upf.edu}} \\ \vspace{0.43cm}}
  {\secondauthor} {Music Institute and Department\\of Electrical Enginering\\Pontificia Universidad\\Católica de Chile\\ %
    {\tt \href{mailto:rcadiz@uc.cl}{rcadiz@uc.cl}}}
  {\thirdauthor} {Institute for Mathematical \\ and Computational Engineering\\Pontificia Universidad\\Católica de Chile\\ %
    {\tt \href{mailto:casinglo@uc.cl}{casinglo@uc.cl}}}
  {\fourthauthor} {Music Technology Group\\Universitat Pompeu Fabra \\ %
    {\tt \href{mailto:frederic.font@upf.edu}{frederic.font@upf.edu}}}
  {\fifthauthor} {Music Technology Group\\Universitat Pompeu Fabra \\ %
    {\tt \href{mailto:xavier.serra@upf.edu}{xavier.serra@upf.edu}}}

\usepackage{xcolor}

\begin{document}

\capstartfalse
\maketitle
\capstarttrue

\begin{abstract}
This paper explores the innovative application of the Fractional Fourier Transform (FrFT) in sound synthesis, highlighting its potential to redefine time-frequency analysis in audio processing. As an extension of the classical Fourier Transform, the FrFT introduces fractional order parameters, enabling a continuous interpolation between time and frequency domains and unlocking unprecedented flexibility in signal manipulation. Crucially, the FrFT also opens the possibility of directly synthesizing sounds in the $\alpha$-domain, providing a unique framework for creating timbral and dynamic characteristics unattainable through conventional methods. This work delves into the mathematical principles of the FrFT, its historical evolution, and its capabilities for synthesizing complex audio textures. Through experimental analyses, we showcase novel sound design techniques, such as $\alpha$-synthesis and $\alpha$-filtering, which leverage the FrFT’s time-frequency rotation properties to produce innovative sonic results. The findings affirm the FrFT’s value as a transformative tool for composers, sound designers, and researchers seeking to push the boundaries of auditory creativity.
\end{abstract}

\section{Introduction}
Sound synthesis has traditionally relied on mathematical tools such as the Fourier Transform (FT) to analyze and generate audio signals. While the FT has served as a cornerstone in audio signal processing, its limitations become apparent when dealing with non-stationary signals or when flexible time-frequency representations are required. The Fractional Fourier Transform (FrFT), a generalization of the FT, addresses these challenges by introducing a fractional order parameter ($\alpha$) that allows continuous interpolation between the time and frequency domains, allowing access to the $\alpha$-domain, a hybrid space in between the traditional time and frequency domains. This capability not only provides enhanced flexibility for signal analysis but also opens the possibility of directly synthesizing and manipulating sounds within the $\alpha$-domain. By exploring the mathematical foundations and applications of the FrFT, this paper investigates its potential as a sound synthesis tool, capable of generating novel timbral and dynamic audio characteristics that extend beyond traditional approaches.

The idea of fractional audio signal processing is not new. In \cite{De_Poli_1984} the authors present a nonlinear sound synthesis technique that utilizes a sinusoidal input signal combined with a rational waveshaping function, allowing for the generation of complex spectral evolutions through the manipulation of various parameters. The term ``fractional'' in this technique is used in a different way, as it refers to the use of this rational waveshaping function, made out of ratios or fractions of polynomials, hence fractional, designed to process a sinusoidal input signal, transforming it into a more complex waveform. This technique is not fractional in the sense of a rotation in the time-frequency domain, but rather in that it uses fractions of polynomials.

The FrFT has also been directly applied to non-synthesis audio signal processing tasks. To enhance the analysis of complex signals like speech, a method based on the Short-Time FrFT (SFrFT) has been proposed \cite{Ram_Palo_Mohanty_2019}. This method improves the time-frequency resolution, allowing for better identification of frequency content in speech, due to the fact that the SFrFT is particularly effective in filtering out unwanted noise and distortion, which leads to improved signal enhancement. Their results indicate that the SFrFT outperforms the conventional FrFT, showing better signal to noise ratio (SNR) and perceptual evaluation of speech quality (PESQ) under various noisy conditions.

Another application for speech signals is based on what is called a fractional spectral subtraction (FSS) for improving noisy speech. The process begins by applying the FrFT to segments of noisy speech samples. After transforming the speech, the method estimates the noise present in the signal and subtracts it from the transformed speech. This subtraction is crucial for isolating the actual speech from the noise. Finally, the enhanced speech is obtained by applying the inverse fractional Fourier transform (IFrFT), which converts the processed data back into a form that can be understood as clear speech. The authors claim that their method can find the best fractional order to separate speech from noise effectively, leading to significant improvements in the quality of the enhanced speech \cite{Wang_Zhang_2005}.

The FrFT has also been used as a more effective compression of sound signals, relying on a more flexible representation compared to the traditional Fourier transform. This flexibility is crucial for effective compression because sound signals can be represented in a way that emphasizes certain frequency components while reducing the representation of others. This selective emphasis helps in minimizing the amount of data needed to represent the signal without significant loss of quality \cite{Aloui_Brahim_2021}.

Despite these applications of the FrFT in the audio domain, we found no direct attempts to synthesize signals in the $\alpha$-domain. Moreover, in \cite{nicol2005timbre}, its use in sound synthesis was dismissed, as it did not offer advantages over the traditional FT for their purposes. In this article, we demonstrate how the FrFT can be directly applied to sound synthesis, enabling the creation of novel timbral possibilities and dynamic sonic textures. By leveraging its unique mathematical property of rotating the time-frequency domain, we aim to showcase the FrFT's potential as a powerful tool for composers and sound designers.

This article is structured as follows: Section 2 provides an overview of the Fractional Fourier Transform (FrFT), including its formal definition, key mathematical properties, and various implementations relevant to audio processing. Section 3 introduces the proposed methods for sound synthesis and processing in the alpha-domain, including $\alpha$-synthesis and $\alpha$-filtering, emphasizing their potential for creative audio applications. Section 4 presents audio and visual examples to illustrate the techniques and explore the unique sonic possibilities enabled by the FrFT. Section 5 discusses the results, highlighting the implications of the FrFT's time-frequency rotation properties on sound synthesis and design. Finally, Section 6 concludes the article by summarizing key findings and outlining potential future directions for research and application.  

\section{Fractional Fourier Transform Overview}
In this section, we formally introduce the Fractional Fourier Transform (FrFT) along with some of its properties, and briefly discuss various implementations. Different implementations can produce significantly different results, so it is important to discuss their differences and explain why we chose one implementation over the others for our examples.
\subsection{Formal Definition and Its Properties}
The roots of the Fractional Fourier Transform trace back to the 1920s, where it was first theorized about by Wiener in \cite{first_frft}. The current definition of the Fractional Fourier Transform however, was introduced in the 1960s by Bargmann \cite{true_frft}, formally defining the Fractional Fourier Transform as an extension of the classical Fourier Transform incorporating a rotation parameter to generalize its behavior in the time-frequency plane. Since then, the FrFT has found applications in optics, quantum mechanics, and, more recently, audio processing.

One of the many equivalent ways of defining the FrFT is as an integral operator parameterized by an angle $\alpha$ that determines the fractional domain of transformation. For a given signal $f(t)$, the FrFT of order $\alpha$ is expressed as:
\[
F_\alpha f (s) = \int_{-\infty}^\infty f(t) K_\alpha(s, t) \, dt,
\]
where $K_\alpha(s, t)$ is the kernel of the FrFT, defined for $\alpha\in(-2,2)\backslash\{0\}$ as
\[
K_\alpha(s,t) = K_\phi \exp\left(i \pi (t^2 \cot \alpha - 2 t s \csc \alpha + s^2 \cot \alpha)\right),
\]
where $\phi = \alpha\pi/2$ and 
\[
K_\phi = \exp\left(-i\left(\frac{\pi\text{sgn}(\phi)}{4} - \frac{\phi}{2}\right)\right)/|\sin(\phi)|^{0.5}.
\]
Additionally, for $\alpha\in \{-2,0,2\}$, one can extend the last definition considering $K_0(s,t) = \delta(s-t)$ and $K_{\pm 2}(s,t) = \delta(s+t)$, and then extend it outside $[-2,2]$ using modular arithmetic modulo 4, that is, forcing $F_\alpha = F_\beta$ if $\alpha\equiv\beta \pmod{4}$.

The property that makes the FrFT a true generalization of the Fourier Transform is that when integer multiples of $\pi/2$ are used for $\alpha$, the FrFT reduces to integer powers of the actual FT, that is,
\begin{equation}\label{eq:integer}
F_{n\cdot \pi/2} = \mathcal{F}^n,\quad \text{for all } n\in\mathbb{Z},
\end{equation}
where $\mathcal{F}$ correspond to the Fourier Transform.

Among many other properties, the FrFT is a linear transform in its argument and a (group) homomorphism from $(\mathbb{R}, +)$ to $(\{F_\alpha\}_{\alpha\in\mathbb{R}}, \circ)$ (also called index additivity \cite{dfrft_alt_1}). These properties can be summarized respectively as:
\begin{align}
F_\alpha(g + h)    &= F_\alpha(g) + F_\alpha(h), \label{eq:prop_linearity} \\
F_{\alpha + \beta} &= F_\alpha \circ F_\beta,    \label{eq:prop_homomorphism}
\end{align}
for all $g,h\in \mathcal{L}^2(\mathbb{R})$ and $\alpha,\beta \in \mathbb{R}$.

Note that this last equation together with property \eqref{eq:integer} tell us that we can easily find the inverse of the FrFT for any value of $\alpha$ by simply changing the sign of the rotation angle. This is because
\[
F_\alpha \circ F_{-\alpha} = F_{\alpha-\alpha} = F_0 = \mathcal{F}^0 = Id.
\]


\subsection{Connections to Linear Canonical Transforms}

The FrFT is a particular case of a symmetric Linear Canonical Transform (LCT)~\cite[Ch.~2]{healy_linear_2016}. Given a \(2\times 2\) symmetric matrix
\[
    {\boldsymbol A} = \begin{bmatrix} a & -b \\ -b & a \end{bmatrix}\qquad a,b\in \mathbb{R}
\]
the symmetric LCT induced by it is the integral transform
\[
    L_A u(f) = \int_{-\infty}^{\infty} K_A(f, t) u(t)\, dt
\]
with kernel
\[
    K_A(f, t) = C_A e^{\pi i (a f^2 - 2b f t + a t^2)}.
\]
The normalization constant is selected so that \(|C_A| = \sqrt{|b|}\). It is apparent that the FrFT of order \(\alpha\) is a symmetric LCT with matrix
\[
    {\boldsymbol A} = \begin{bmatrix} \cot \alpha & -\csc\alpha \\ -\csc\alpha & \cot\alpha \end{bmatrix}.
\]
Any symmetric LCT is also unitary, satisfying Plancherel's identity~\cite[Sec.~2.2]{healy_linear_2016}
\[
    \int_{-\infty}^{\infty} u(t) v(t)^*\, dt = \int_{-\infty}^{\infty} L_A u(f) L_A v(f)^*\, df.
\]
A remarkable property of a symmetric LCT is its action on the Wigner distribution of a signal \(u\)
\[
    W_u(t, f) = \int_{-\infty}^{\infty} u\left(t + \frac{1}{2}s\right) u\left(t - \frac{1}{2}s\right)^* e^{-2\pi i f s}\, ds.
\]
The Wigner distribution is real-valued, but it may take negative values. It is typically interpreted as a distribution of the time-frequency content of \(u\) on the time-frequency plane. It can be proven that~\cite[Sec.~2.3]{healy_linear_2016}
\[
    W_{L_A u}(t, f) \cong W_u\left(\frac{a}{b} f - \frac{1}{b} t,\, \frac{b^2 - a^2}{b} f + \frac{a}{b} t\right).
\]
In other words, a symmetric LCT rotates and scales the time-frequency distribution of \(u\). In the particular case of the FrFT of order \(\alpha\) the above becomes
\[
    W_{L_A u}(t, f)  \cong W_u (f\cos\alpha - t\sin\alpha, f\sin\alpha + t\cos\alpha).
\]
Consequently, the FrFT of order \(\alpha\) is a rotation in the time-frequency plane. This property is particularly useful when sonically interpreting the results given by the FrFT, and we created some examples to illustrate this phenomenon. Such examples can be found in subsection \ref{subsec:examples_rotation}.


\subsection{Implementations}\label{subsec:implementations}
Over the years, the FrFT has been implemented digitally in various ways. In \cite{fast_frft}, the first two implementations of this transform were presented. These implementations approximated the FrFT of a continuous function by decimating and finding relations between the computation of the FrFT and the FFT. This resulted in very fast implementations (specifically, $o(N\log(N))$), but did not ensure the homomorphism property \eqref{eq:prop_homomorphism}.

Subsequently, the discussion shifted towards defining a discrete version of the FrFT (see \cite{dfrft_og}, \cite{dfrft_alt_1} and \cite{dfrft_alt_2}). Although one could use the earlier implementations as a definition for the discrete case, this was not considered due to the lack of the homomorphism property \ref{eq:prop_homomorphism}. Each proposed definition had its pros and cons, and most of them assured the properties that made the FrFT interesting, but they did not manage to offer a computation method fast enough for real-time audio purposes.

Different definitions, conventions, and implementations have been used across various fields over the years. However, the references cited earlier appear to have had the most significant impact on the signal processing field. In recent years, several Python implementations, some even tailored for use in the deep learning domain, have been proposed (see \cite{frft_pytorch}). Due to these advancements and their speed, for this research, we decided to consider the implementations of \cite{fast_frft} given in \cite{frft_pytorch}.

\section{Fractional Fourier Transform Synthesis and Processing Methods}
In this section, we explore sound synthesis and manipulation methods strongly based on the FrFT. While we believe much more can be done, we have chosen to keep this simple, as our aim is to consolidate the foundations of the creative use of the FrFT in audio.

Before discussing any methods, some clarifications and decisions need to be made. As previously discussed, the FrFT implementations in Subsection \ref{subsec:implementations} can be used to manipulate any type of discrete complex signal. In particular, any type of sound signal can be transformed at a glance with any angle. The result will most likely be a complex signal, so decisions must be made to make these results audible. Additionally, one must decide whether to transform the entire signal at once or to process it in windows, then overlap and add the resulting frames to generate a final signal.

The methods we propose will use only the real or imaginary part of the transformed signals. Moreover, we will focus on using low values of $\alpha$, typically less than $0.5$, and manipulating the window size as needed. The values of $\alpha$ have been shown to be of critical importance for transforming audio signals. In \cite{Jin_Chen_Wei_Xia_2012}, the authors analyze two main aspects of the FrFT when applied to audio data: the sensitivity of the rotation factor and the diffusion of the transform. The rotation factor is a key parameter in the FrFT that affects how signals are transformed. Their results indicate that the sensitivity of the rotation factor is significant, meaning that even small changes in the rotation factor can lead to noticeable differences in the output signal. Additionally, the paper discusses diffusion, which refers to how changes in the fractional Fourier domain data can affect the ability to accurately restore the original signal. If any element of the fractional Fourier domain changes by more than 0.2 times the original data, the inverse transform will completely fail to recover the original signal. 

To understand the decision to use low values of $\alpha$, it is also important to recall that the FrFT can be regarded as a rotation by an angle of $\alpha\pi/2$ in the time-frequency domain. This implies that using values where $\alpha > 0.5$ will transform the audio input into a representation closer to the frequency domain than the time domain. As expected, the spectrum of a real signal is not of particular sonic interest.

\begin{figure}
    \centering
    \includegraphics[width=1\linewidth]{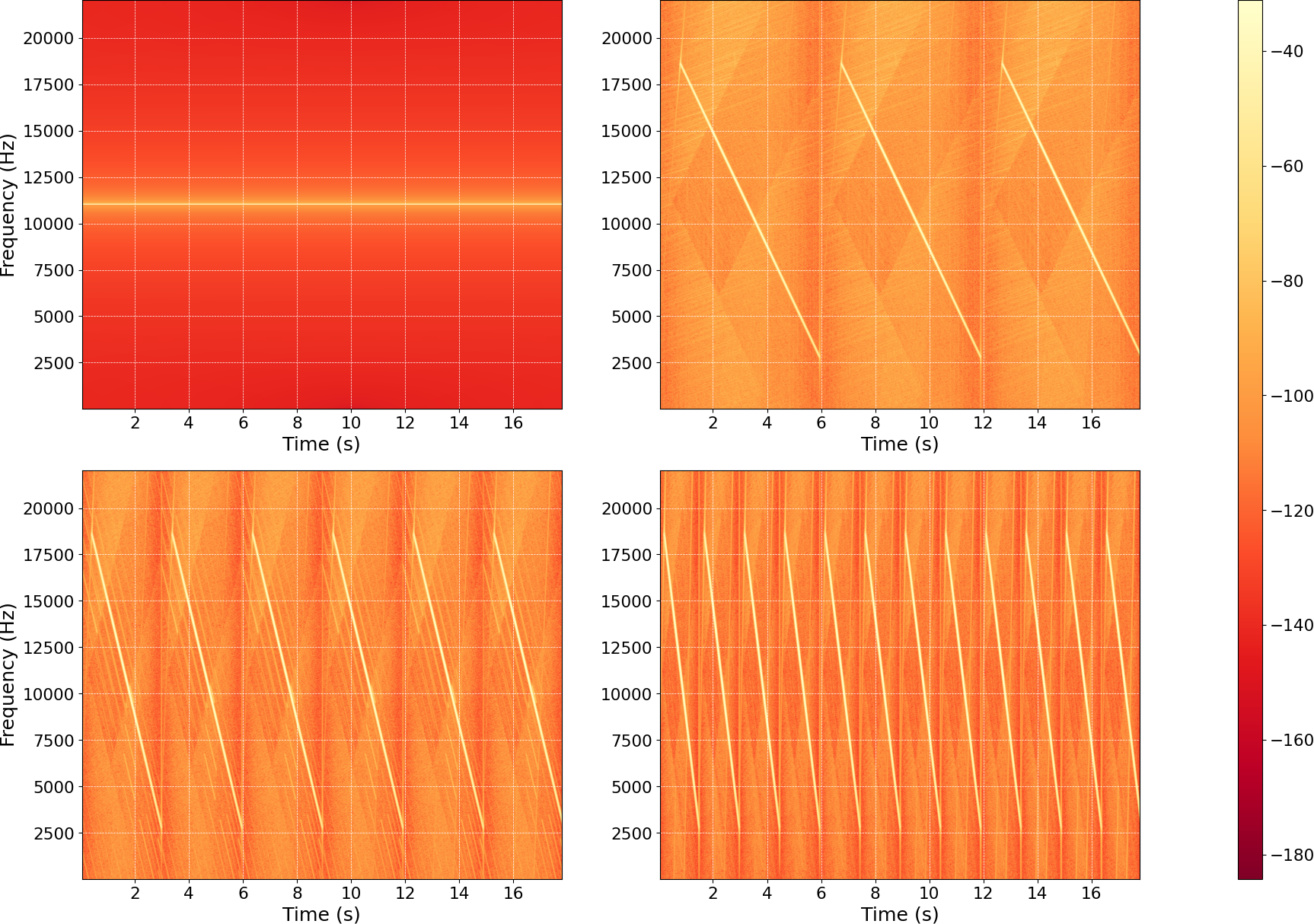}
    \caption{Spectrogram of the FrFT of a sinusoid with a frequency of 11025 Hz using window sizes corresponding to 1.48 s, 2.97 s, 5.94 s, and 11.88 s (the entire signal).}
    \label{fig:rotated_spectrograms}
\end{figure}

Regarding window size, it is crucial to understand that merging a sequence of rotated spectrograms will generally produce very different results. Figure \ref{fig:rotated_spectrograms} illustrates this concept.

\subsection{Method 1: $\alpha$-Synthesis}\label{subsec:method_1}
The FrFT can transform any type of sound, and while it might generate interesting sounds, such a transform is extremely complex. Consequently, it would be challenging to feel in control when using it broadly. This problem, however, diminishes when the input sound is fixed to one whose spectrogram is fully understood. By doing so, and recalling the spectrogram rotation property of the FrFT, one can anticipate the results generated by the transform and gain some control over it.

Based on this, we introduce the $\alpha$-Synthesis method as the application of the FrFT, with the previously mentioned limitations, on pure sinusoids. Sinusoids are clear examples of sounds with simple spectrograms that are easy to understand, allowing the use of the FrFT to generate complex yet intelligible sounds.

\subsection{Method 2: $\alpha$-Filtering}\label{subsec:method_2}
It is well known that the Fourier Transform can be used to create filters in the frequency domain through simple multiplication. Additionally, the Convolution Theorem allows this concept to be translated to the time domain via convolutions. This property led to the development of Finite Impulse Response (FIR) and Infinite Impulse Response (IIR) filters, which can be designed to be quite accurate and, in some cases, computed more efficiently. Unfortunately, there is no known analog to the Convolution Theorem for the FrFT, so filtering a signal using this transform necessitates pure multiplication.

Given this, what exactly is filtering using the FrFT? The human auditory system has a well-studied sensitivity to frequency, making frequency domain filters a reasonable and understandable process in audio contexts. On the other hand, the FrFT transforms signals into the $\alpha$-domain, a mixture of frequency and time. To our knowledge, the human auditory system does not have a well-established understanding of this domain. Specifically, in the frequency domain, each band corresponds to a pure sinusoid in the time domain. However, a band in the $\alpha$-domain corresponds to a signal that heavily depends on the value of $\alpha$, and to our understanding, no studies have explored human perception of such signals.

For our purposes, we propose the $\alpha$-Filtering method as a natural generalization of frequency domain filtering. Specifically, a signal is first transformed into the $\alpha$-domain using the FrFT, then multiplied with a kernel, and finally transformed back to the time domain using the corresponding inverse FrFT. Similar to the frequency domain, one can define $\alpha$-low pass, $\alpha$-band pass, and $\alpha$-high pass filters using appropriate kernels. However, it must be understood that "low," "band," and "high" in this context will generally not correspond to frequency. For example, an $\alpha$-low pass filtered signal might have richer high-frequency content than its $\alpha$-high pass counterpart.

\section{Audio and Visual Examples}
Accompanying this article, we have created several videos and sound examples\footnote{Videos, sound examples and code are available in this articles webpage \href{https://cordutie.github.io/frft_sound_synthesis/}{cordutie.github.io/frft\_sound\_synthesis/}.} showcasing a series of sound examples built using the techniques discussed herein. A brief explanation of some of the visual and sonic examples is provided in this section, but more examples can be found on this article's webpage.

\subsection{Time-Frequency Domain Rotation}\label{subsec:examples_rotation}
The rotation property of the FrFT is well-studied; however, we decided to include an example in this research as it is fundamental to understanding its implications on sounds.

\begin{figure}[H]
    \centering
    \includegraphics[width=1\linewidth]{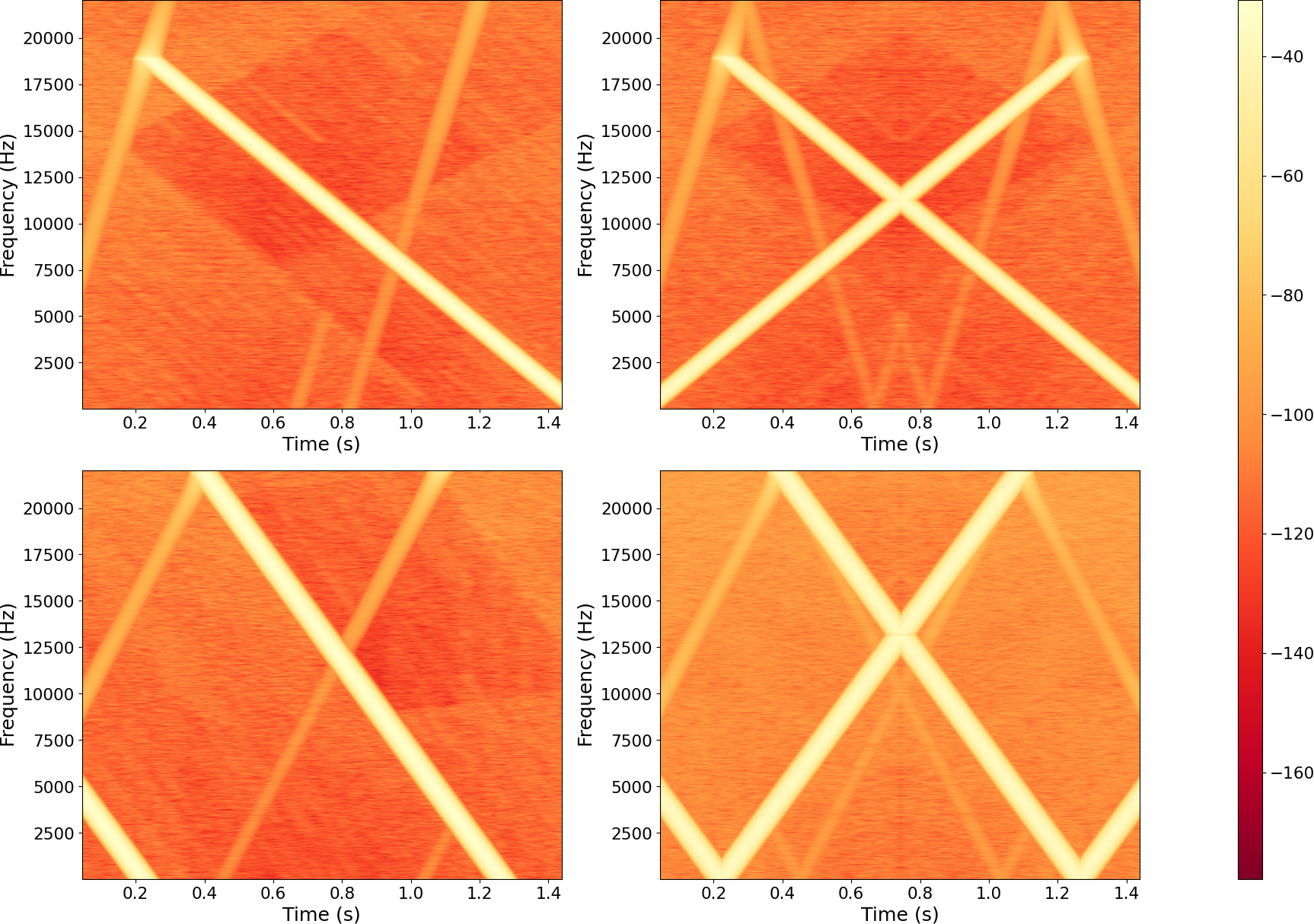}
    \caption{Spectrograms of the full FrFT versus its real part of a sinusoid of frequency $10025$ Hz. Left column corresponds to the full FrFt while right column corresponds to the real part only. First row corresponds to $\alpha=0.3$ and second corresponds to $\alpha=0.45$.}
    \label{fig:FrFT_complex_vs_real}
\end{figure}

For this example, we considered a sinusoid with a frequency of 
$10025$ Hz (exactly in the middle of the sampleable frequency domain) and computed both the spectrogram of its FrFT and the real part of its FrFT for values of $\alpha$ ranging from $0$ to $1$. Note that the FrFT in this case is computed over the entire signal. Each video correspond to one of the transformation mentioned before and a figure comparing both is added in Figure \ref{fig:FrFT_complex_vs_real},

\subsection{Sound Example Group 1: $\alpha$-Synthesis}\label{subsec:examples_synthesis}
In our first group of sound examples, we used Method 1 (see \ref{subsec:method_1}) to generate audio from the real part of the FrFT of several sinusoids with various window lengths.

Specifically, one second of audio is generated from a sinusoid for each frequency in $\{55, 220, 880\}$ Hz. These sounds are then processed using the FrFT with window sizes of $\{0.5, 1\}$ seconds, hop sizes equal to half of the window sizes, and angles in $\{0, 0.01, 0.05, 0.1, 0.25, 0.5\}$. Each video corresponds to one frequency and one window size setting, and all transforms are displayed sequentially with angles increasing as mentioned. Spectrum and spectrogram representations are available in separate videos.

\subsection{Sound Example Group 2: $\alpha$-Synthesis + Angle Manipulation}\label{subsec:examples_synthesis_2}
In this group of sound examples, we use Method 1 by computing the FrFT of several sinusoids while manipulating the value of $\alpha$ throughout the transform.

Specifically, one second of audio is generated from a sinusoid for each frequency in $\{55, 220, 880\}$ Hz. These sounds are then processed using the FrFT with window sizes of approximately $0.046, 0.092, 0.18$, and $0.32$ seconds, with hop sizes equal to half of the respective window sizes, and values of $\alpha$ increasing linearly from $0$ to $0.5$. Each video corresponds to one frequency, with all transforms displayed sequentially as the window sizes increase according to the values mentioned. Spectrum and spectrogram representations are provided in separate videos.

\subsection{Sound Example Group 3: $\alpha$-Filtering + Band Manipulation}\label{subsec:examples_filtering_bands}
In our third group of sound examples, we used Method 2 (see \ref{subsec:method_2}) to filter two sinusoids with an $\alpha$-band pass filter. These $\alpha$-band pass kernels are created by multiplying in the $\alpha$-domain with the spectrum of an impulse response of the form
\begin{equation}\label{eq:impuls_response}
IR(t) = \exp(-0.5((tb)^2)) \cos(2\pi c t),
\end{equation}
where $t$ corresponds to time, $b$ to the bandwidth of the filter, and $c$ to the center frequency of the filter. The filters are applied over time using fixed values of $\alpha$ and bandwidth $b$, while varying the center frequencies $c$.

Specifically, two seconds of two sinusoids at frequencies $220$ Hz and $3520$ Hz are generated and $\alpha$-band pass filtered using the spectrum of impulse responses in the form of \eqref{eq:impuls_response} with $b=1$, $c$ increasing exponentially (base $2$) from $100$ to $10000$, and values of $\alpha$ in $\{0.01, 0.05, 0.1, 0.25, 0.5\}$. The filtering is done using window sizes of approximately $0.19$ and $0.38$ seconds, with hop sizes equal to half of the respective window sizes. Each video corresponds to a specific frequency and window size setting, with all transforms displayed sequentially as the angles increase according to the specified values.
\section{Results and Discussion}
Our examples concerning the rotation property of the FrFT confirm that this property is clearly satisfied when considering the full (complex) FrFT of the signal. More significantly for this research, they also demonstrate that when only the real part is considered, the spectrogram appears to exhibit a reflection against a horizontal line corresponding to the fundamental frequency of the original signal. This results in an additional chirp in the spectrum, moving in the opposite direction to the original. Furthermore, once the lines intersect the boundaries determined by time, new reflections against these bounds emerge.

Examples from Group 1 illustrate that when the FrFT is applied to large windows, it can generate distinct streams of chirps traversing the time-frequency domain in parallel lines, along with other lines that appear to be symmetries about the horizontal line representing the fundamental frequency of the input, as well as reflections on the boundaries of the space.

Examples from Group 2 demonstrate that this approach shares similarities with certain aspects of traditional FM synthesis, particularly when using a high modulation index. In this case, the increasing fractional rotation angle $\alpha$ resembles the effect of increasing both the carrier and modulation frequencies in FM synthesis, leading to a progressively richer harmonic spectrum. The gradual manipulation of $\alpha$ introduces dynamic spectral changes that parallel the evolving complexity in FM synthesis, offering a comparable yet distinct framework for achieving intricate timbral variations.

Finally, examples from Group 3 indicate that bands in the $\alpha$-domain, despite being theoretically distinct (as they form a basis), produce sounds that are sonically similar. This is evident as shifting the center frequency of the $\alpha$-band pass filter from $c=100$ to $c=10000$ does not cause a dramatic change in the sound, unlike what is typically expected in a frequency-domain filter. Moreover, though this may be anecdotal, our observations suggest a weak inverse relationship between the bands of the $\alpha$-domain and the frequency domain. Specifically, $\alpha$-filtering in lower bands tends to produce sounds with a richer high-frequency spectrum compared to their higher band counterparts.

\section{Conclusion}
This study examines the application of the Fractional Fourier Transform (FrFT) in sound synthesis, introducing methods that exploit its time-frequency rotation property to synthesize and process sounds in the $\alpha$-domain. The proposed approaches, $\alpha$-synthesis and $\alpha$-filtering, demonstrate how the FrFT can be used to generate complex and timbrally diverse audio signals. These methods provide a novel framework for exploring the interplay between time and frequency domains, offering composers and sound designers a broader set of tools for sound manipulation.

The results of this research highlight the potential of the FrFT for creative audio applications. The $\alpha$-synthesis method illustrates how simple input signals, such as sinusoids, can be transformed into complex outputs through fractional domain transformations, with the rotation angle serving as a key parameter for control. Similarly, the $\alpha$-filtering technique generalizes traditional filtering concepts to the alpha-domain, enabling new forms of signal processing that combine time- and frequency-domain characteristics. These findings underscore the FrFT’s capacity to facilitate innovative approaches to sound design and synthesis.

Future work could further investigate the perceptual implications of alpha-domain transformations and refine these methods for integration into real-time audio systems. Additionally, exploring the use of the FrFT in contexts such as interactive systems, spatial audio, and adaptive music could provide valuable insights into its broader applicability. By establishing a foundation for the use of FrFT in sound synthesis, this study contributes to expanding the possibilities for both scientific research and artistic practice in the field of audio signal processing.

\begin{acknowledgments}
This research was supported by ANID Fondecyt Regular Grant \#1230926, ANID Anillo ATE220041, Government of Chile, and the project "IA y Música: Cátedra en Inteligencia Artificial y Música (TSI-100929-2023-1)" funded by the "Secretaría de Estado de Digitalización e Inteligencia Artificial and the Unión Europea-Next Generation EU". We would also like to thank Diego Vera for his contributions to an early version of this project during his undergraduate research.
\end{acknowledgments} 

\bibliography{auxiliar/references.bib}

\end{document}